\documentclass[letter]{aa} 
\usepackage{graphicx}
\usepackage{txfonts}
\usepackage{natbib}
\bibpunct{(}{)}{;}{a}{}{,}
\usepackage{epsfig}
\begin{document}
\title{Phase resolved spectroscopy of the accreting millisecond X-ray
  pulsar SAX J1808.4$-$3658 during the 2008 outburst\thanks{Based on
    observations made with ESO Telescopes at the Paranal Observatory
    under programme ID 281.D-5060(A).}}


\author{R. Cornelisse$^1$, P. D'Avanzo$^2$, T. Mu\~noz-Darias$^{1}$,
  S. Campana$^2$, J. Casares$^{1}$, P.A. Charles$^{3,4}$, D.
  Steeghs$^{5,6}$, G. Israel$^7$, L. Stella$^7$ }

   \offprints{corneli@iac.es}

   \institute{$^1$ Instituto de Astrofisica de Canarias, Calle Via Lactea S/N, E-3805 La Laguna, Spain\\
$^2$ INAF- Osservatorio Astronomico di Brera, via E. Bianchi 46, 23807 Merate, Italy\\
$^3$ South Africa Astronomical Observatory, P.O. Box 9, Observatory 7935, South Africa\\
$^4$ School of Physics and Astronomy, University of Southampton, Highfield, Southampton SO17 1BJ, UK\\
$^5$ Department of Physics, University of Warwick, Coventry, CV4 7AL, UK\\
$^6$ Harvard-Smithsonian Center for Astrophysics, 60 Garden Street, Cambridge,
MA 02138, USA\\
$^7$ INAF-Osservatorio Astronomico di Roma, Via Frascati 33, I-00040
Monteporzio Catone (Rome), Italy\\}

   \date{Received; accepted}

 
  \abstract 
{} 
{We obtained phase-resolved
  spectroscopy of the accreting millisecond X-ray pulsar SAX J1808.4-3658
  during its outburst in 2008 to find a signature of the
  donor star, constrain its radial velocity semi-amplitude ($K_2$), 
  and derive estimates on the pulsar mass.  } 
{Using Doppler images of the Bowen region we find a significant
  ($\ge$8$\sigma$) compact spot at a position where the donor star is
  expected. If this is a signature of the donor star, we measure
  $K_{em}$=248$\pm$20 km s$^{-1}$ (1$\sigma$ confidence) which
  represents a strict lower limit to $K_2$. Also, the Doppler
  map of He\,II $\lambda4686$ shows the characteristic signature of
  the accretion disk, and there is a hint of enhanced emission that may
  be a result of tidal distortions in the accretion disk that are
  expected in very low mass ratio interacting binaries.}
{The lower-limit on $K_2$ leads to a lower-limit on the mass function
of $f$($M_1$)$\ge$0.10$M_\odot$. Applying the maximum $K$-correction
gives 228$<$$K_2$$<$322 km s$^{-1}$ and a mass ratio of
0.051$<$$q$$<$0.072.}
{Despite the limited S/N of the data we were able to detect a
  signature of the donor star in SAX\,J1808.4$-$3658, although future
  observations during a new outburst are still warranted to confirm
  this. If the derived $K_{em}$ is correct, the largest uncertainty in
  the determination of the mass of the neutron star in
  SAX\,J1808.4-3658 using dynamical studies lies with the poorly known
  inclination.}

\titlerunning{Spectroscopy of SAX\,J1808.4$-$3658}
\authorrunning{Cornelisse et~al.}

   \maketitle
%

\section{Introduction}

Low-Mass X-ray Binaries (LMXBs) are systems in which a compact object
(a neutron star or a black hole) is accreting matter, via Roche lobe
overflow, from a low mass ($<$1$M_{\odot}$) star. Some LMXBs exhibit
sporadic outburst activity but for most of their time remain in a
state of low-level activity (White et~al. 1984); we will refer to
these systems as transients.  In April 1998, a coherent 2.49 ms X-ray
pulsation was discovered with the {\it Rossi X-ray Timing Explorer}
({\it RXTE}) satellite in the transient SAX J1808.4$-$3658 (Wijnands
\& van der Klis 1998, Chakrabarty \& Morgan 1998). This was the first
detection of an Accreting Millisecond X-ray Pulsar (AMXP). Seven more
of these systems have been discovered since then; all these systems
are transients, have orbital periods in the range between 40 min and
4.3 hr and spin frequencies from 1.7 to 5.4 ms.  These findings
directly confirmed evolutionary models that link the neutron stars of
LMXBs to those of millisecond radio pulsars via the spinning up of
the neutron star due to accretion during their LMXB phase (e.g.
Wijnands 2006).

To date, six episodes of activity were detected from SAX
J1808.4$-$3658 with a 2-3 year recurrence cycle. During the 1998
outburst a detailed analysis of the coherent timing behaviour showed
that the neutron star was in a tight binary system with a 2.01 hr
orbital period (Chakrabarty \& Morgan 1998; Hartman et~al. 2008).  The
mass function derived from X--ray data ($4\times 10^{-5} M_{\odot}$)
and the requirement that the companion fills its Roche lobe led to the
conclusion that it must be a rather low mass star, possibly a brown
dwarf (Chakrabarty \& Morgan 1998; Bildsten \& Chakrabarty 2001).  

\begin{figure*}[t]\begin{center}
\psfig{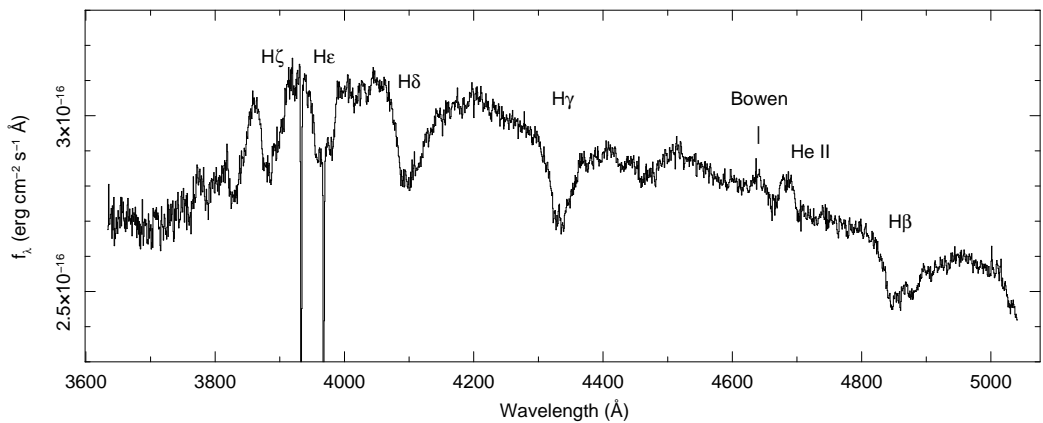}
\caption{Average flux calibrated spectrum of SAX\,J1808.4$-$3658. We
have labelled the most important features that are present. The strong
absorption features around 3940 and 3980 \AA~ are due to diffuse interstellar
bands.
\label{average}}
\end{center}\end{figure*}

We present here optical spectroscopy of SAX J1808.4$-$3658 obtained
during the 2008 outburst.  One of the key issues in dynamical studies
is to measure the radial velocity of the companion star ($K_2$) and
use this value to constrain the optical mass function of the system.
During quiescence, these goals can be achieved by tracing the
absorption features originating in the photosphere of the companion
star. However, the intrinsic faintness of the low mass companion stars
in AMXPs makes such an analysis very difficult.  To overcome this
problem, Steeghs \& Casares (2002) have shown that during phases of
high mass accretion rates, Bowen blend lines emitted by the irradiated
face of the companion star can be used. A precise measurement of $K_2$
represents the only way to determine the optical mass function of the
system and ultimately constrain the neutron star mass.

\section{Observations and data reduction}

Optical spectroscopic observations of SAX J1808.4$-$3658 were carried
out on 27 September 2008 with the ESO Very large Telescope (VLT), using the
1200B grism on FORS1 with a slit width of 0.7 arcsec. We obtained 16
spectra of 360 seconds integration each, which corresponds to one
orbital period. The seeing during the observations was in the range
$0.8\arcsec{}-1.1\arcsec{}$. 

Image reduction was carried out following standard procedures:
subtraction of an averaged bias frame, division by a normalised flat
frame. The extraction of the spectra was performed with the
ESO-MIDAS\footnote{http://www.eso.org/projects/esomidas/} software
package. Wavelength and flux calibration of the spectra were achieved
using a helium-argon lamp and by observing spectrophotometric standard
stars. Our final reduced spectra have a wavelength range from 3600-5000 \AA,
a dispersion of 0.72 \AA~pixel$^{-1}$ and resolution of R=2200.
Cross correlation of the spectral lines and the Doppler
tomograms were obtained using the MOLLY and DOPPLER packages developed
by Tom
Marsh\footnote{http://deneb.astro.warwick.ac.uk/phsaap/software/}.

\section{Data analysis}

We present the average spectrum of SAX\,J1808.4$-$3658 in
Fig.\,\ref{average}. The spectrum is dominated by strong Balmer lines
in absorption, which are due to the optically thick disk in the high
state and are a typical signature of a low-to-intermediate inclination
system. The He\,II $\lambda$4686 and the Bowen complex (at
$\lambda$$\lambda$4630-4660) are also clearly detected as emission
features, and we have indicated the most important lines.

For other bright LMXBs, narrow components in the Bowen emission have
been reported that are thought to arise on the irradiated surface of
the donor star ( e.g. Steeghs \& Casares 2002; Casares et~al.
  2006; Cornelisse et~al.  2007; see also Cornelisse et~al. 2008 for
  an overview), and here we attempt to find similar features in
SAX\,J1808.4$-$3658. To calculate the orbital phase for each spectrum
we used the recent ephemeris by Hartman et~al. (2008), but added 0.25
orbital phase to their phase zero so that it represents inferior
conjunction of the secondary.  Unfortunately, the individual spectra
do not have sufficient S/N to identify the narrow features, and we
must resort to the technique of Doppler tomography (Marsh \& Horne
1988).  This technique uses all the spectra simultaneously to probe
the structure of the accretion disk and identify compact emission
features from specific locations in the binary system. However, for
this technique to work, an estimate of the systemic velocity of the
system, $\gamma$, is crucial.  We do want to note that by changing the
phase zero of the accurate Hartman et~al.  (2008) ephemeris any
potential donor star feature in the map must now lie along the
positive y-axis in the Doppler tomogram. Thus finding a significant
feature there will give strong support to an emission site located on
the irradiated donor star.

\begin{figure}\begin{center}
\psfig{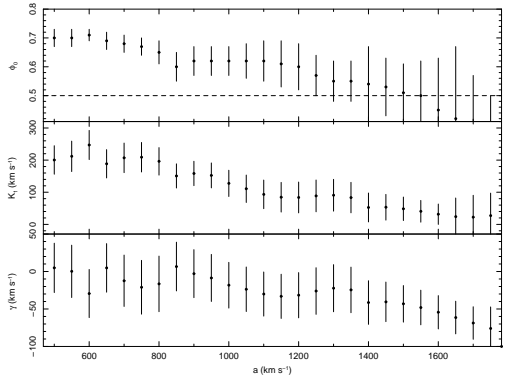}
\caption{The derived fit parameters of the radial velocity curve of the
He\,II $\lambda4686$ emission as trace by a double Gaussian with separation 
$a$. The free parameters are the orbital phase zero, $\phi_0$, 
compared to the ephemeris of Hartman et~al. (2008), 
its radial velocity, $K_1$, and its systemic velocity, $\gamma$. The dotted
line in the top panel indicates the expected phase zero for the compact 
object.
\label{gaus}}
\end{center}\end{figure}

To find $\gamma$ we started by applying the double-Gaussian technique
of Schneider \& Young (1980) to He\,II $\lambda$4686. Since the wings
of the emission line should trace the inner-accretion disk, it should
not only give us an estimate of the already known radial velocity of
the compact object (Chakrabarty \& Morgan 1998), but also $\gamma$.
Using a Gaussian band pass with FWHM of 400 km s$^{-1}$ and
separations between 500 and 1800 km s$^{-1}$ in steps of 50 km
s$^{-1}$, we find that between a separation of 1400 to 1600 km
s$^{-1}$ our values for $K_{1}$ are close to the one obtained by
Chakrabarty \& Morgan (1998), while at larger separations we are
reaching the end of the emission line (see Fig.\,\ref{gaus}). Also in
this range, our fitted orbital phase zero ($\phi_0$) is close to 0.5,
further suggesting that we are tracing the radial velocities of a
region close to the neutron star, while the systemic velocity is
stable around -50 km s$^{-1}$. Despite the large errors on our fits,
we do think this test already gives a good first estimate of $\gamma$
around -50 km s$^{-1}$.

\begin{figure}[t]
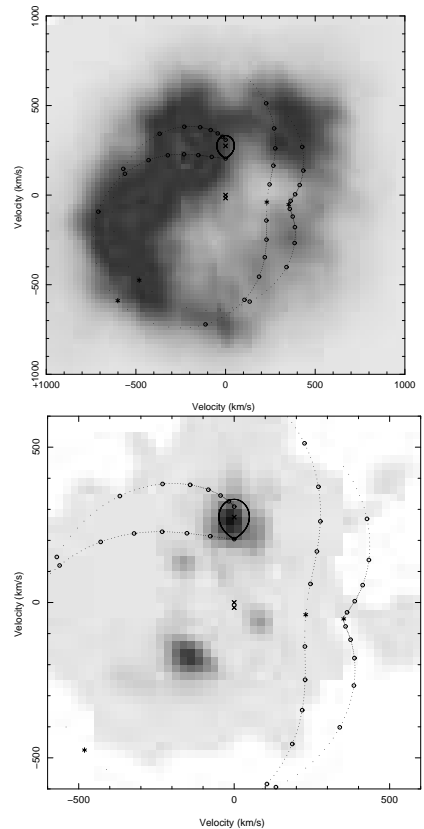
\begin{center}
\psfig{figure=SAX1808_hedop.ps,width=5.5cm, angle=-90}
\psfig{figure=SAX1808_dopp_bw.ps,width=5.5cm, angle=-90}
\caption{Doppler maps of He\,II $\lambda$4686 (top) and
the Bowen complex (bottom). Indicated on both maps is 
the Roche lobe, gas stream leaving the L$_1$ point, and the Keplerian
velocity along the stream for the assumptions $q$=0.059 and
$K_2$=275 km s$^{-1}$.
\label{doppler}}
\end{center}\end{figure}

Another test to obtain $\gamma$ is to create Doppler maps for He\,II
$\lambda4686$. Contrary to the Bowen region (which is very complex due
to the presence of many different lines), He\,II is a single line that
is usually a good tracer of the accretion disk (see e.g. Cornelisse
et~al. 2007; Casares et~al. 2006). We searched for $\gamma$ between
-160 and 0 km s$^{-1}$ in steps of 20 km s$^{-1}$, and all the maps
show the expected accretion disk structure (see Fig.\,\ref{doppler}).
We must unfortunately conclude that He\,II is not very sensitive to
$\gamma$, and only suggests a range between -160 and 0 km s$^{-1}$. We
note that the maps are dominated by an emission feature in the
top-left quarter of the map, which we interpret as the gas stream
impact point. We also note that further downstream there is enhanced
emission, that might be due to matter streaming along the edge of the
disk as was observed in for example EXO\,0748$-$676 (Pearson et~al.
2006). Finally we note that there is some enhanced emission in the
top-right corner which might be due to strong tidal interaction (see
below).

\begin{figure}\begin{center}
\psfig{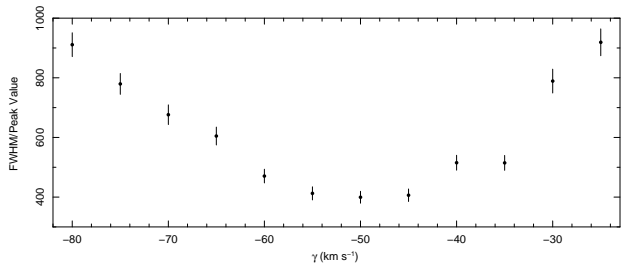}
\caption{The ratio of the FWHM over the peak value for the compact spot
in the Bowen map (using only N\,III $\lambda$4661) as a function of the
systemic velocity ($\gamma$).
\label{gamma}}
\end{center}\end{figure}

Cornelisse et~al. (2008) have shown that the strongest narrow
component in the Bowen emission is usually N\,III $\lambda$4641. Our
next step was therefore to create Doppler maps of the Bowen region for
$\gamma$ between 0 and -120 km s$^{-1}$ (again in steps of 20 km
s$^{-1}$) including only this line. Only when $\gamma$ was between -80
and -20 km s$^{-1}$ was a clear spot present and centred on the $x$=0
axis.  For this range we estimated the peak value and FWHM of the spot
as a function of $\gamma$, and in Fig.\,\ref{gamma} we show how the
ratio of these values change. Around $\gamma$=-50$\pm$15 km s$^{-1}$
(1$\sigma$ confidence) the FWHM/peak value reaches a minimum
suggesting that here the spot is most compact, and this is the value
we adopt for the systemic velocity. Furthermore, we also noted that in
the range from -35 to -65 km s$^{-1}$ the velocity centroid of the
spot was stable between $V_y$=240-260 km s$^{-1}$.

To decrease the noise present in the Bowen Doppler map we included the
most important other lines that are most often present in other LMXBs
(N\,III $\lambda$4634 and C\,III $\lambda$4647/4650), and present this
map in Fig.\,\ref{doppler}. To estimate the significance of the
compact spot we measured the standard deviation of the brightness of
the pixels in the background. We find that the central pixels of the
compact spot are 18$\sigma$ above the background, and even 8$\sigma$
above the second most prominent feature in the map, namely the one at
(-150,-200) in the Bowen map of Fig.\,\ref{doppler} (for which it is
unclear if it is real or an artifact of the tomogram). This strongly
suggest that the compact spot is real.

Finally, to optimise our estimate for $K_{em}$ we created average
spectra in the rest frame of the donor star, changing $K_{em}$ in
steps of 2 km s$^{-1}$ within our error range.  We find that N\,III
$\lambda$4640 is most pronounced for $K_{em}$=248$\pm$20 km s$^{-1}$
(1$\sigma$ confidence), and adopt this as our final value, and in
Fig.\,\ref{donor} show the final average spectrum in the rest-frame of
the donor star. We do note that this value is smaller than the
$\simeq$300 km s$^{-1}$ obtained by Ramsay et~al. (2008) from the same
dataset, since they provide no errors we cannot tell if this
difference is significant. However, as stated above, the location of
the spot remains within the quoted error range for a range of assumed
$\gamma$ velocities. To search for variability in the Bowen lines as a
function of orbital phase, we created an average spectrum in the
rest-frame of the donor using only spectra taken between orbital
phases 0.25 and 0.75, and another corresponding to phases 0.75 and
1.25. The strength of N\,III $\lambda$4640 did not change between
these spectra, but it is unclear if this is real (suggesting that the
inclination is low) or due to the limited S/N of the dataset.

\section{Discussion}

\begin{figure}\begin{center}
\psfig{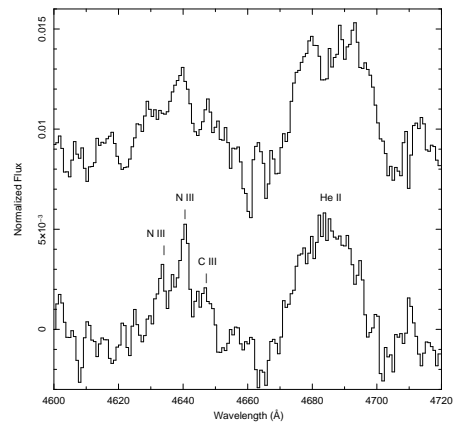}
\caption{Blow-up of the Bowen region for the average spectrum (top) and
the average in the rest-frame of the donor star (bottom). The bottom
spectrum clearly shows the narrow components. Indicated are 
the most important N\,III ($\lambda$4634/4640) and C\, III ($\lambda$4647) 
lines. 
\label{donor}}
\end{center}\end{figure}

We have presented phase-resolved spectroscopy of the accreting
millisecond X-ray pulsar SAX\,J1808.4$-$3658, and detected a compact
feature in the Doppler map of the Bowen complex. Although the S/N of
the data is limited, this spot is the most stable and significant
feature for a large range of $\gamma$ velocities. Furthermore, thanks
to the very accurate ephemeris (Hartman et~al. 2008), the spot is at a
position where the donor star is expected, and we conclude that it is
real.  Therefore, following detections of a donor star signature in
other X-ray binaries (e.g. Steeghs \& Casares 2002; Casares et~al.
2006; Cornelisse et~al. 2007; see Cornelisse et~al.  2008 for an
overview), we also identify this feature as being produced on the
irradiated surface of the donor star.  We note that in
Fig.\,\ref{donor} most peaks in the Bowen region appear to line up
with known N\,III and C\,III lines (e.g. Steeghs \& Casares 2002) when
using our derived values for $\gamma$ and $K_{em}$.

Since the donor star surface must have a lower velocity than the
centre of mass, the observed $K_{em}$=248$\pm$20 km s$^{-1}$
(1$\sigma$ confidence) is a lower limit on the true $K_2$ velocity.
However, it still gives us a strict lower limit on the mass function
of $f(M)$=$M_1$sin$^3i$/(1+$q$)$^2$$\ge$0.10$M_\odot$, where $q$ is
the binary mass ratio $M_2$/$M_1$ and $i$ the inclination of the
system. We can further constrain the mass function by applying the
so-called $K$-correction (Mu\~noz-Darias et~al. 2005), and using the
fact that $K_1$=16.32 km s$^{-1}$ (Chakrabarty \& Morgan 1998). The
largest $K$-correction possible is when we assume that there is no
accretion disk and almost all radiation is produced in the L$_1$
point.  Applying the polynomials by Mu\~noz-Darias et~al. (2005) for
$K_{em}$=248$\pm$20 km s$^{-1}$ gives $K_2$=299$\pm$23 km s$^{-1}$,
which should be independent of the inclination of the system.

This gives conservative estimates of 228$<$$K_2$$<$322 km s$^{-1}$ and
0.051$<$$q$$<$0.072, which we used to create the Roche lobe and gas
streams on the Bowen Doppler map in Fig.\,\ref{doppler}.  We do note
that our obtained mass ratio is rather extreme, and would suggest that
tidal interaction in SAX\,J1808.4$-$3658 is important enough to
produce a precessing accretion disk and thereby a superhump (see e.g.
O'Donoghue \& Charles 1996). This might be the explanation of the
enhanced emission in the top-right quarter of the He\,II map
(Fig.\,\ref{doppler}/top), which was for example also observed
in LMC\,X-2 (Cornelisse et~al. 2007). Such a superhump should be
moving through the accretion disk on the precession time scale, and
therefore changes position in the Doppler map over time.
Unfortunately, since we only have one orbit of data we cannot test
this, and future observations will be needed to see if the spot is
long-lived and moves, in order to confirm the presence of the
superhump.

Despite our large range of $K_2$ we nonetheless will review the
implications for the mass of the neutron star. First of all we can
improve our estimate of $K_2$ by taking into account the results by
Meyer \& Meyer-Hofmeister (1982). They analysed the effects of X-ray
heating on the accretion disk and found that there is a minimal disc
opening angle of $\simeq$6$^\circ$, even in the absence of
irradiation. Using this value to estimate the $K$-correction, the
polynomials by Mu\~noz-Darias (2005) suggest that $K_2$$\le$310 km
s$^{-1}$, still comparable to the maximum correction possible. Deloye
et~al. (2008) used photometry of SAX\,J1808.4$-$3658 in quiescence to
constrain the inclination between 36 and 67 degrees.  Using these
extreme values for their inclination and 228$<$$K_2$$<$322 km
s$^{-1}$, leads to a neutron star mass between 0.15 and 1.58$M_\odot$.
Although these values are not very constraining for the neutron star
mass, they do favour a mass near the canonical value rather than a
massive neutron star. We also note that our values are smaller than
the $>$1.8$M_\odot$ estimated by Deloye et~al. (2008) (for a 10\%
error in the distance estimate), but since this corresponds to a
1.7$\sigma$ difference (only taking into account our errors), we
conclude that this disagreement is marginal.

\section{Conclusions}

The observations presented here provide evidence for the detection of
the signature of the irradiated donor star in SAX\,J1808.4$-$3658.
Clearly a similar experiment, but at higher S/N and spectral
resolution, must be carried out again during a future outburst to
obtain more than a single orbital period of data and resolve the
narrow lines. This will not only allow us to unambiguously claim the
presence of narrow components in the spectra of SAX\,J1808.4$-$3658,
but also measure the rotational broadening of the narrow components to
further constrain $K_2$ via the relation in Wade \& Horne (2003).  With
these data we have shown a promising way forward to constrain $K_2$
and in combination with more quiescent data, we should be able to
better constrain the inclination and thereby obtain the mass of the
neutron star in SAX\,J1808.4$-$3658.

%

\begin{acknowledgements}
  We cordially thank the director of the European Southern Observatory
  for granting Director's Discretionary Time (ID 281.D-5060(A). We
  like to thank the referee Craig Heinke,for the careful and helpful
  comments which have improved this paper. RC acknowledges a Ramon y
  Cajal fellowship (RYC-2007-01046). RC acknowledges Katrien
  Uytterhoeven for useful discussion on different analysis techniques.
  PDA and SC thank S. Covino for useful discussions.  DS acknowledges
  a STFC Advanced Fellowship. JC acknowledges support by the Spanish
  MCYT GRANT AYA2007-66887.  Partially funded by the Spanish MEC under
  the Consolider-Ingenio 2010 Program grant CSD2006-00070: First
  Science with the GTC.
\end{acknowledgements}

\end{document}